\documentclass[12pt]{article}
\usepackage[totalheight = 23cm, totalwidth = 17cm]{geometry}
\usepackage{graphicx, subfigure}
\newcommand{\nin}{\noindent}
\newcommand{\be}{\begin{equation}}
\newcommand{\ee}{\end{equation}}
\newcommand{\bea}{\begin{eqnarray}}
\newcommand{\eea}{\end{eqnarray}}

\newcommand{\nn}{\nonumber\\}

\newcommand{\ol}{\overline}
\usepackage{amsmath}
\usepackage{amssymb}

\begin{document}

\begin{center}

{\bf{\Large Quantization leading to a natural \\ flattening of the
axion potential}}

\vspace{0.5cm}

J. Alexandre\footnote{jean.alexandre@kcl.ac.uk} and D. Tanner\footnote{dylan.tanner@kcl.ac.uk}

King's College London, Department of Physics, WC2R 2LS, UK

\vspace{0.5cm}

{\bf Abstract}

\end{center}

\nin  Starting from the general cosine form for the axion effective
potential, we quantize the axion and show that the
result is described by a naturally flat potential, if interactions
with other particles
are not considered. This feature therefore restores the would-be
Goldstone-boson nature of the axion, and we calculate the
corresponding vacuum energy density, which does not need to be too large by orders of 
magnitude compared to Dark Energy.

\vspace{1cm}

{\it Introduction.}
It is known that a quantized scalar theory must be described by a convex one-particle-irreducible (1PI) potential \cite{convex},
as a consequence of the so-called
spinodal instability, where the restoration force on field fluctuations vanishes.
In the situation of a bare potential presenting a concave region, as a Mexican-hat potential,
the corresponding 1PI potential can be obtained by a procedure similar to the Maxwell construction in Thermodynamics
\cite{Maxwell}. This property is valid for a self-interacting scalar field only, and therefore does not apply to
a situation where the scalar also interacts with other degrees of freedom, such as the Higgs boson in the Standard Model.

An example of the self-interacting case is provided by the Sine-Gordon model, in 1+1 dimensions, where Wilsonian
renormalization studies have been done, dealing with the corresponding spinodal instability \cite{Sine-Gordon}.
In these studies, the authors consider the coarse grained potential, which interpolates the bare potential and the 1PI potential,
when the cut off of the theory decreases. They find
that, between two minima of the cosine bare potential, the coarse grained potential flattens
when the cut off decreases, and becomes flat in the IR limit.
We note here that this flatness holds in a wider area
than the one where the bare potential is concave, and the flat 1PI potential joins the minima of the bare potential.
As a consequence,
starting from a cosine bare potential, the only way for the 1PI potential to be convex is to be flat.
Other studies involving a bare cosine potential, and the corresponding flattening due to spinodal instability,
have been made in the context of quitessence \cite{quintessence}.  The authors consider a
description based on spinodal decomposition, using a potential function of two variables: the mean scalar field
and its fluctuation condensate, in the framework of a Hartree approximation.

We consider here a cosine potential, in 3+1 dimensions, which is
characteristic of axions, and we show how a flat potential indeed
arises. As a consequence, starting with a cosine bare potential, the
axion self-interactions gradually flatten the potential, and we are
left with a vacuum energy only, which we calculate. A novel feature
of the present study is that we quantize the axion degrees of
freedom as compared to regarding the axion as a classical field in
conventional cosine-based expressions for the potential. Using this
fully quantized treatment we derive an expression for the
self-interacting axion vacuum energy and we see that in some heavy
axion models, this vacuum energy is not necessarily orders of
magnitude larger than observed and estimated values for Dark Energy.
Coupling the axion to other particles of the Standard Model, though,
can prevent the flattening of the potential, and we discuss a
specific example, where the axion is coupled to fermions. We derive
conditions to avoid the spinodal instability, such that no
flattening of the axion potential occurs.

This article is structured as follows.
We first review the quantization of the axion model, in order to define our notations, and we show that the corresponding
1PI potential must be convex. We then derive
an exact self consistent equation for this 1PI potential, in the form of a differential Schwinger-Dyson equation,
which is non-perturbative. A flat 1PI potential is indeed a solution of
this equation, and we calculate the corresponding vacuum energy density.
Finally, We discuss the coupling of the axion to other particles.

\vspace{0.5cm}

{\it The axion and its quantization.} In the original paper
\cite{PQ}, the axion is introduced to build a CP conserving theory,
from a model with massive fermions coupled to a non-Abelian gauge
field, where a CP violating term arises at one-loop due to
non-perturbative QCD instanton effects, \cite{'t Hooft}.  This
original model was refined following experimental and cosmological
evidence and \cite{J Kim Axion Summary} proposes a more general
Lagrangian with three potentially non-zero coupling constants from
which the original work \cite{PQ}, or invisible axion models such as
\cite{KSVZ}, can be constructed.  In this paper we use
the term axion in this general sense, defined as initially arising
as a phase - a Goldstone mode - of a complex scalar field, $\Phi$
which acquires a vacuum expectation value $\langle \Phi\rangle =
fe^{i\theta}$ due to spontaneous breaking of a global $U(1)$
symmetry at a scale $f$. Explicit symmetry breaking occurs at some
scale $\mu$, with $\mu < f$ at which point the axion can acquire
mass. Motivated by the need to solve the CP problem, we require that
the axion model possess a non-zero anomalous gluon coupling term
$\theta G\widetilde{G}$. References
 \cite{J Kim Axion Summary}, \cite{Witten}, and \cite{QCD Vacuum and Axions}
 quote the commonly used general expression for the axion potential as
 $K (1- \cos \theta)$, where $K$ has mass dimension 4. 
In arriving at this expression the initial step is
integrating over the $QCD$ degrees of freedom. The total partition function is
\be\label {QCD quant}
Z=\int{\cal D}[\phi,\Phi]\exp\left(-S[\phi,\Phi]\right)=\int{\cal D}[\theta]\exp\left(-S_\theta\right),
\ee
where $\phi$ represents the $QCD$ degrees of freedom and $\Phi=\rho e^{i\theta}$.
$S_\theta$ defines then the effective action for the axion, 
which we will quantize. The calculation of $S_\theta$ is necessarily approximate, and an
analysis of the form of the axion potential in light of new
understandings of the $QCD$ vacuum, using supersymmetric gauge and
brane theories, is presented in \cite{QCD Vacuum and Axions}. The authors note that the $K(1- \cos \theta)$ form
for the potential of the $QCD$ axion needs not be limited to one cosine
dependence, but may have higher harmonics and in general is a smooth
periodic function of $\theta$ with period $2 \pi$. They conclude
that with axion models using light quarks $m_q \ll \Lambda_{QCD}$
one arrives at a cut-off independent potential of the form 
$K(1-\cos \theta)$.

Motivated by these results and the general form of the Lagrangian presented in
\cite{J Kim Axion Summary}, we use as our general form for the axion
action as follows.
\be\label{model} 
S_{\theta}=\int d^4x\left\lbrace
\frac{f^2}{2}\partial_\mu\theta\partial^\mu\theta +\sum_{n=1}^\infty
a_n(1 -(\cos\theta)^n)\right\}, 
\ee
where $a_n$ are coefficients to be determined (units quartic in
mass). The action (\ref{model}) can be understood as a consequence of the upgrading
of the phase of $\langle\Phi\rangle$ to a dynamical field, that we wish to quantized. An example of  
quantization of the axion is done in \cite{Bose}, where the authors consider quantum fluctuations of the axion,
leading to a Bose-Einstein condensate which could play the role of cold Dark Matter. 

In terms of the canonically
normalized field $\phi=f\theta$, this theory is not renormalizable, and we will need to keep the cut off
$\Lambda$ throughout the study,
which will be considered a parameter of the theory, as well as $f$. 
Also, by definition of the cut off, we will always assume
that $f\le\Lambda$.  (For completeness, we also assume that $\Lambda_{QCD}\le \Lambda$.)

Quantization of the model (\ref{model}) is based on the partition function.
\be\label{Z} 
Z[j]=e^{-W[j]}=\int{\cal D}[\theta]\exp\left(-S_\theta-\int d^4x~ j\theta\right), 
\ee 
from which the classical field is defined as 
\be
\theta_{cl}\equiv\frac{\delta W}{\delta j}=\left<\theta\right>, 
\ee
where 
\be 
<\cdots>=\frac{1}{Z}\int{\cal
D}[\theta](\cdots)\exp\left(-S_\theta-\int d^4x~ j\theta\right).
\ee 
A second functional derivative with respect to the source gives
\be\label{d2W} 
-\frac{\delta^2 W}{\delta j\delta j}=
\left<\theta\theta\right>-\theta_{cl}\theta_{cl}, 
\ee 
which is the variance of the set of variables $\theta(x)$, therefore positive, such that 
\be\label{negative} 
\frac{\delta^2 W}{\delta j\delta j}\le0,
\ee 
which will be used to show the convexity of the 1PI potential.
The proper graph generator functional $\Gamma[\theta_{cl}]$ is
defined as the Legendre transform of the connected graphs generator
functional $W[j]$: 
\be 
\Gamma[\theta_{cl}]=W[j]-\int
d^4x~j~\theta_{cl}, 
\ee 
where the source $j$ is understood as a functional of $\theta_{cl}$, and we have then 
\be\label{d2G}
\frac{\delta\Gamma}{\delta\theta_{cl}}=-j~~~~~~~~\mbox{and}~~~~~~~~
\frac{\delta^2\Gamma}{\delta\theta_{cl}\delta\theta_{cl}}=-\left(\frac{\delta^2W}{\delta
j\delta j}\right)^{-1}. 
\ee 
The convexity of the 1PI potential $U$
follows directly from the previous properties \cite{convex}. $U$ is
the non derivative part of $\Gamma$: if $\theta_0$ is a constant
classical field, $\Gamma[\theta_0]=\int d^4xU(\theta_0)$. The
corresponding source, denoted $j_0$, is also constant, such that the
relation between second functional derivative in eqs.(\ref{d2G})
gives then, after integration over space time, 
\be\label{d2U}
\frac{d^2U}{d\theta_0^2}=-\int d^4x\left( \frac{\delta^2W}{\delta
j\delta j}\right)_{j_0}^{-1}, 
\ee 
such that, taking the inequality
(\ref{negative}) into account, we have, for any value of $\theta_0$,
\be 
\frac{d^2U}{d\theta_0^2}\ge0, 
\ee 
and the potential is
necessarily convex. As a consequence, if the bare potential is
concave, at least in a certain range of field values, quantum
corrections necessarily contain tree-level contributions, in order
to compensate the concave features of the
bare potential and make the 1PI potential convex.\\
If the axion is coupled to other degrees of freedom, then
$\delta^2W$ has to be understood as a matrix, with rows and columns
corresponding to the different degrees of freedom. The second
derivative of the 1PI potential is then an element of the inverse
matrix $(\delta W)^{-1}$, which is not necessarily positive. In this
latter situation,
concave features of the bare potential can be stable, and quantum corrections are perturbative only.\\
Note that, if one considers a periodic axion field, which is
restricted to the interval $[0;2\pi]$ in the path integral
(\ref{Z}), the latter should be defined with 
\be 
\int{\cal D}[\theta]\equiv\Pi_{x}\int_0^{2\pi}d\theta(x), 
\ee 
but the
derivation of the property (\ref{negative}) would not change.
Indeed, this inequality was not obtained by any explicit path
integration, but corresponds to a generic property of $Z[j]$. For
the same reason, the existence of non-trivial vacua and topological
features in the classical theory does not alter the property
(\ref{negative}): all the corresponding information is contained in
the path integral. Only a saddle point
approximation would have to take into account these topological features, but no approximation or expansion is made here.\\
Finally, by construction, the effective potential must be differentiable, since it is
obtained by integrating over the original field $\theta$. The only way for the effective potential to be convex, differentiable
and periodic is to be flat: we will see now that this solution is indeed satisfied.

\vspace{0.5cm}

{\it Self consistent equation for the 1PI potential.} Our approach
consists in studying the evolution of the proper graph generator
functional $\Gamma$ with the amplitude of the decay constant $f$.
This functional approach is inspired from the original version given
in \cite{original}, and will lead us to an exact partial
differential equation for the 1PI potential $U$. In a first stage,
this procedure can be seen as a mathematical tool, to obtain a
non-perturbative self-consistent equation for $U$. Nevertheless, the
physical meaning of the approach is the following. Suppose we start
from $f>>m$, the interaction is then negligible compared to the
kinetic term, and the theory is almost free: quantum fluctuations
are ``frozen''. As $f$ decreases, quantum fluctuations start to
dress the system, which will eventually describes the proper quantum
theory when $f$ reaches its physical value. This argument will be
used to define the boundary condition necessary to solve the self-consistent differential equation that we now derive.\\
From now on, we represent a derivative with respect to $f$ with a dot. We have then
\be
\dot\Gamma=\dot W+\int d^4x\frac{\delta W}{\delta j}\partial_fj-\int d^4x~\partial_f j~\theta_{cl}
=\dot W,
\ee
where the latter derivative can be obtained from the bar action (\ref{model}), such that
\be\label{dotG}
\dot W=f\int d^4x\left<\partial^\mu\theta\partial_\mu\theta\right>~.
\ee
In what follows, we omit the subscript {\it cl} for the classical field.
From eq.(\ref{d2W}), we obtain then
\be
\dot\Gamma=f\int d^4xd^4y~\delta^{(4)}(x-y)\frac{\partial}{\partial x^\mu}\frac{\partial}{\partial y_\mu}
\left\lbrace\theta_x\theta_y-\frac{\delta^2W}{\delta j_x\delta j_y}  \right\rbrace, 
\ee
and, using eq.(\ref{d2G}), we finally find that
the exact evolution equation of $\Gamma$ with $f$ is
\be\label{evoleq}
\dot\Gamma=f\int d^4x~\partial^\mu\theta\partial_\mu\theta
+f\int d^4xd^4y~\delta^{(4)}(x-y)\frac{\partial}{\partial x^\mu}\frac{\partial}{\partial y_\mu}
\left(\frac{\delta^2\Gamma}{\delta\theta_x\delta\theta_y}\right)^{-1},
\ee
where the trace contains all the quantum corrections, and is regularized by the cut off $\Lambda$.
We stress that this equation is exact: it is a self-consistent equation for
$\Gamma$, which is independent of any perturbative expansion.\\
Next step is to consider an ansatz for $\Gamma$, and we take here
\be\label{ansatz} 
\Gamma=\int d^4x\left(
\frac{f^2}{2}\partial^\mu\theta\partial_\mu\theta+U(\theta)\right),
\ee 
which allows an $f$-dependent 1PI potential $U(\theta)$. It is then sufficient to consider a constant axion configuration: 
the evolution of the effective potential is then given by $\dot\Gamma={\cal V}\dot U(\theta)$, 
where ${\cal V}$ is the volume of space time. We have thus
\be
\frac{\partial}{\partial x^\mu}\frac{\partial}{\partial y_\mu}
\left(\frac{\delta^2\Gamma}{\delta\theta_x\delta\theta_y}\right)^{-1}
=\int\frac{d^4p}{(2\pi)^4}\frac{d^4q}{(2\pi)^4}\frac{-p^\mu q_\mu}{f^2p^2+U''}\delta^{(4)}(p+q)e^{-ipx-iqy},
\ee
such that, taking into account ${\cal V}=\delta^{(4)}(p=0)$,
\be
\int d^4xd^4y~\delta^{(4)}(x-y)
\frac{\partial}{\partial x^\mu}\frac{\partial}{\partial y_\mu}
\left(\frac{\delta^2\Gamma}{\delta\theta_x\delta\theta_y}\right)^{-1}
={\cal V}\int\frac{d^4p}{(2\pi)^4}\frac{p^2}{f^2p^2+U''(\theta)},
\ee
where a prime denotes a derivative with respect to the axion.
The integration over $p$ leads then to the following non-perturbative and
self-consistent equation for the 1PI potential 
\be\label{evolU} 
\dot U(\theta)=\frac{1}{16\pi^2f}\left[\frac{\Lambda^4}{2}-\frac{\Lambda^2}{f^2}U''(\theta)
+\frac{1}{f^4}[U''(\theta)]^2\ln\left(1+\frac{f^2\Lambda^2}{U''(\theta)}\right)
\right] , 
\ee 
where $\Lambda$ is the UV cut off used to regularize
the theory. There are several solutions to the non-linear partial
differential equation (\ref{evolU}), but, as expected from the
spinodal instability discussed above, a flat potential $U_{vac}$ is
indeed a solution, and satisfies 
\be\label{dotU} 
\dot U_{vac}=\frac{\Lambda^4}{32\pi^2f} \ee We consider some key points at this stage:
\begin{itemize}

\item The flattening of the potential cannot be found with a perturbative expansion,
but only with a non-perturbative approach: quantum fluctuations are so dominant in such a system,
that they wipe off the classical non-convex features of the bare potential;

\item The cosines in the bare potential do not appear anywhere in the evolution equation (\ref{evolU}), 
because the latter is a self-consistent equation for the dressed potential. The cosines
of the bare potential will play a role as a boundary condition used to solve
this differential equation;

\item The axion, originally a Goldstone boson, and acquiring mass after integration of $QCD$ degrees of freedom,
eventually indeed becomes massless, after quantization of the model (\ref{model});

\item The flat potential $U_{vac}$ is not related to some vacuum expectation value of the axion field, and therefore
does not need to satisfy any specific transformation under parity;

\item The equation (\ref{dotU}) can be written as a time-energy uncertainty relation, $f\tau=1$, where
\be
\tau=32\pi^2\frac{\dot U_{vac}}{\Lambda^4}.
\ee
Hence, although we are dealing with a field theory at equilibrium, one might
interpret $\tau$ as the time that quantum fluctuations need to flatten the potential.
This interpretation gives a physical meaning to the quantity $\dot U_{vac}/\Lambda^4$.

\end{itemize}

\vspace{0.5cm}

{\it Boundary condition.} The integration of eq.(\ref{dotU})
necessitates a boundary condition $f=\Lambda\Rightarrow
U=U_\Lambda$, such that 
\be\label{flatpot}
U_{vac}=U_\Lambda+\frac{\Lambda^4}{32\pi^2}\ln\left(
\frac{f}{\Lambda}\right). 
\ee 
In order to determine $U_\Lambda$, we
use the following argument. When $f=\Lambda$, the bare potential
expressed in terms of the canonically normalized field
$\phi=\Lambda\theta$ is 
\be\label{bare} 
V(\phi) =\sum_{n=1}^\infty a_n(1 - (\cos\theta)^n)=\frac{1}{2}\frac{m^4}{\Lambda^2}\phi^2
+{\cal O}\left(\frac{\phi}{\Lambda}\right)^4, 
\ee
where 
\be
m^4\equiv\sum_{n=1}^\infty na_n. 
\ee
Assuming $m<<\Lambda$,
the interactions are negligible compared to the kinetic and mass
energies: the corresponding dressed flat potential $U_\Lambda$ is
then given by its one-loop expression, which is exact if the bare
potential (\ref{bare}) is considered quadratic. The flattening of
the potential is a tree level effect, and therefore of zeroth order
in $\hbar$: without further corrections, this would lead to a flat
potential interpolating the minima of the classical potential, at
zero energy. The one-loop correction, of first order in $\hbar$, is
then identified with $U_\Lambda$, 
\bea
U_\Lambda&=&\frac{1}{2}\mbox{Tr}\left\lbrace \ln\left[(-\partial^2+V'')\delta^{(4)}(x-y)\right]\right\rbrace \\
&=&\frac{\Lambda^4}{64\pi^2}\left[ \left(
1-\frac{m^8}{\Lambda^8}\right) \ln\left(
1+\frac{\Lambda^4}{m^4}\right)
-\frac{1}{2}+\frac{m^4}{\Lambda^4}\right] \nn &\simeq&
\frac{\Lambda^4}{16\pi^2}\ln\left(
\frac{\Lambda}{m}\right),\nonumber 
\eea 
such that the flat potential (\ref{flatpot}) is finally 
\be\label{vacpot} 
U_{vac}\simeq
\frac{\Lambda^4}{32\pi^2}\ln\left( \frac{f\Lambda}{m^2}\right). 
\ee
To be consistent with the interpretation in terms of vacuum energy,
we need $U_{vac}\ge0$, such that the following constraint should be
satisfied 
\be 
f\Lambda\ge m^2, 
\ee 
which is a property common to
every axion model. Note that the naive estimate of the vacuum energy
density $\rho_{vac}$, based on the bare cosine potential, is
\be\label{naive}
\rho_{vac}=\frac{1}{2}\int\frac{d^3k}{(2\pi)^3}\sqrt{k^2+\frac{m^4}{f^2}}\simeq\frac{\Lambda^4}{16\pi^2}.
\ee 
Although the ratio $f\Lambda/m^2$ is in principle large, the
result (\ref{vacpot}) is actually of the same order of magnitude
than the estimate (\ref{naive}), because of the logarithm of this
ratio. However, for heavy axions \cite{heavy}, the ratio
$f\Lambda/m^2$ can actually be of order 1 given $m = m(f)$, see for
example \cite{axion cosmology} where $m$ is quadratic in $f$.  In
such a situation, the vacuum energy (\ref{vacpot}) can be much
smaller than the naive estimate (\ref{naive}), such that the
contribution of the axion vacuum energy is not necessarily orders of
magnitude larger than observed and estimated values for Dark Energy.
We note attempts to link the axion with Dark Energy \cite{axion dark energy}
and that our result (\ref{vacpot}) can be a model-independent starting
point for such links.

\vspace{0.5cm}

{\it Interactions with other particles and discussion.} One could
argue that, because of the flatness of the potential, the initial
motivation for the axion is lost, since no restoring force traps the
axion in some vacuum expectation value, which compensates CP
violation. In order to prevent this, we now consider a more
realistic scenario where the axion is coupled to leptons with mass
$\mu$, via the term \cite{different} \be gf\theta\ol\psi\gamma^5\psi
\ee where $g$ is a dimensionless coupling, and we will discuss the
conditions under which the spinodal instability can be avoided.
Integrating out leptons, we obtain the effective potential (before
quantizing of the axion):
\be\label{effpot}
V_{eff}(\theta)=
\sum_{n=1}^\infty a_n(1-(\cos\theta)^n)
-\frac{1}{2}\int\frac{d^4p}{(2\pi)^4}\ln\left(\frac{p^2+\mu^2+g^2f^2\theta^2}{p^2+\mu^2}\right),
\ee 
where the integral is regulated by the fermion cut off $M$. The
spinodal instability, present with the cosine potential only, is
avoided if the second derivative of the effective potential
(\ref{effpot}) is positive, such that the axion does not need to
become massless. The original cosine potential is concave at
$\theta=\pi$ for example, where we evaluate the second derivative of
the effective potential (\ref{effpot}): \bea\label{second}
\frac{d^2V_{eff}}{d\theta^2}(\pi)
&=&-m^4-\frac{g^2f^2M^2}{16\pi^2}~\frac{M^2+\mu^2+3g^2f^2\pi^2}{M^2+\mu^2+g^2f^2\pi^2}\nn
&&~~~~~~+\frac{g^2f^2}{16\pi^2}(\mu^2+3g^2f^2\pi^2)\ln\left(
1+\frac{M^2}{\mu^2+g^2f^2\pi^2}\right) \eea and we are left with the
following situations (we neglect the mass $\mu$):
\begin{itemize}

\item if $M>>gf$, the second derivative (\ref{second}) is
\be\label{V2nd}
V_{eff}''\simeq-m^4-\frac{g^2f^2M^2}{16\pi^2}<0,
\ee
and the spinodal instability is not avoided. This is expected for a large fermion cut off, since fermions then
have a negative contribution to the axion effective potential, and do not improve convexity;

\item if $M\simeq gf\pi$, then
\be
V_{eff}''\simeq-m^4+c^4g^4f^4,
\ee
where $c=[(3\ln2-2)/16]^{1/4}\simeq1/4$, and the spinodal instability can thus be avoided if $gf$ is larger than
the threshold $\simeq4m$, which is realistic, given the phenomenological constraints \cite{different}.

\end{itemize}
Note that, if the axion interacts with a scalar degree of freedom
$\Phi$, via the coupling $lf^2\theta^2\Phi^2/4$, where $l$ is a
dimensionless coupling, the correction (\ref{V2nd}) becomes \be
V_{eff}''\simeq-m^4+\frac{lf^2 M^2}{32\pi^2}, \ee and can be
positive, if $l$ is not too small, in which case no concave feature
needs to be compensated by quantum fluctuations. This discussion can
be extended to other interactions with the Standard Model, for
example by coupling the axion to the photon via the interaction \be
\frac{h}{4}\theta
F^{\mu\nu}F^{\rho\sigma}\epsilon_{\mu\nu\rho\sigma}, \ee where $h$
is a dimensionless coupling. 

The essential results from this study can be summarised as follows.
\begin{itemize}

\item When one takes into account quantum fluctuations of the axion,
spinodal instabilities result in a flattening of the effective
potential and a massless axion.  We derive a model-independent form
for the resulting vacuum energy which need not be orders of
magnitude higher than estimates for dark energy.

\item In this quantized axion model, when axion couplings to other particles are considered, we
find that spinodal instabilities can be avoided, allowing the axion
to acquire mass as is the case with conventional axion models.

\end{itemize}

\vspace{0.5cm}

\nin{\bf Acknowledgments} We would like to thank Malcolm Fairbairn for useful comments.

\end{document}